\documentclass[pra,aps,superscriptaddress,twocolumn,floats,nofootinbib,showpacs]{revtex4}
\usepackage{amsmath,amssymb,graphicx}

\newcommand{\dx}{\partial_x}

\newcommand{\PRL}[3]{Phys. Rev. Lett.~\textbf{#1}, #2 (#3)}
\newcommand{\PRB}[3]{Phys. Rev. B~\textbf{#1}, #2 (#3)}
\newcommand{\PRA}[3]{Phys. Rev. A~\textbf{#1}, #2 (#3)}
\newcommand{\PR}[3]{Phys. Rev.~\textbf{#1}, #2 (#3)}
\newcommand{\RMP}[3]{Rev. Mod. Phys.~\textbf{#1}, #2 (#3)}
\newcommand{\Science}[3]{Science~\textbf{#1}, #2 (#3)}

\newcommand{\JETPLett}[3]{JETP Lett.~\textbf{#1}, #2 (#3)}
\newcommand{\JETP}[3]{Sov. Phys. JETP~\textbf{#1}, #2 (#3)}
\newcommand{\ZhETF}[3]{Zh. Eksp. Teor. Fiz.~\textbf{#1}, #2 (#3)}
\newcommand{\ZhETFPis}[3]{Pis'ma Zh. Eksp. Teor. Fiz.~\textbf{#1}, #2 (#3)}

\newcommand{\etal}{\textit{et al.}}

\begin{document}

\title{Photo-Solitonic Effect}

\author{M. Khodas\footnote{Current address: Brookhaven National Laboratory,
Upton, NY 11974}}
\affiliation{William I. Fine Theoretical Physics
Institute,
University of Minnesota, Minneapolis, MN 55455} \affiliation{ School
of Physics and Astronomy, University of Minnesota, Minneapolis, MN
55455}
\author{A. Kamenev}
\affiliation{ School of Physics and Astronomy, University of
Minnesota, Minneapolis, MN 55455}
\author{L. I. Glazman}
 \affiliation{Department of Physics, Yale
University, P.O. Box 208120,
  New Haven, CT 06520-8120}

\begin{abstract}
  We show that dark solitons in 1D Bose liquids may be created by
  absorption of a single quanta of an external ac field, in a close
  analogy with the Einstein's photoelectric effect.  Similarly to the
  von Lenard's experiment with photoexcited electrons, the external
  field's photon energy $\hbar\omega$ should exceed a certain
  threshold.  In our case the latter is given by the soliton energy
  $\varepsilon_s(\hbar q)$ with the momentum $\hbar q$, where $q$ is
  photon's wavenumber.  We find the probability of soliton creation to
  have a power-law dependence on the frequency detuning
  $\omega-\varepsilon_s/\hbar$.  This dependence is a signature of the
  quantum nature of the absorption process and the orthogonality
  catastrophe phenomenon associated with it.
\end{abstract}

 \maketitle

\section{Introduction}

The existence of dark solitons (DS) is among the most spectacular
manifestations of the role played by {\em weak} inter-particle
interactions in 1D cold atomic gases \cite{SolitonObs}. Such
solitons are {\em macroscopically} large areas of partially, or even
completely depleted gas, which propagate coherently without any
dispersion.  It is natural to interpret these objects as localized
solutions of the {\em semi-classical} Gross--Pitaevskii equation
\cite{SolitonTheory}. Correspondingly, the means to create DS,
employed so far, required a macroscopic classical perturbation
applied to the atomic cloud. An example of the latter is the phase
imprinting technique \cite{Imprinting}, where a finite fraction of
the 1D atomic cloud is subject to an external potential for a
certain time. Once the potential is switched off and the gas is
allowed to evolve, the DS is formed around the place with the
maximal gradient of the potential.

Drawing an analogy with the electronic field-emission from a metal: a
pulse of a strong external electric field may lead to creation of free
electrons outside of the metal surface.  It is well-known, however,
from the time of von Lenard and Einstein \cite{Lenard,Einstein} that
this is not the only way to excite electrons. Indeed, a weak ac
field results in a photoelectric current, as long as the energy of its
quanta exceeds the threshold given by the work function of the metal.
The difference between the field-emission and the photoelectric
effects is that the latter essentially utilizes the quantum nature of
the electromagnetic radiation.  Is there an analog of the
photoelectric effect for excitation of DS? Namely can the DS be
created by a weak ac radiation with the frequency exceeding a certain
threshold?

In the framework of the Gross-Pitaevskii equation the answer on these
questions is negative. Indeed, a weak external field may lead to
excitation of the linear waves, if its wavenumber and frequency
satisfy Bogoliubov dispersion relation, but not to creation of DS.
However, treating the Bose liquid beyond the semiclassical
Gross-Pitaevskii approximation reveals that creation of DS in response
on an absorption of a single quanta with an above-the-threshold energy
is actually possible. In analogy with the photoelectric effect we call
this phenomenon the {\em photo-solitonic effect}.  The threshold
energy is given by the energy of DS $\varepsilon_s(p)$ with the
momentum $p=\hbar q$, where $q$ is the photon wavenumber. Creation of
DS requires an ac field with frequency $\hbar\omega>\varepsilon_s(p)$.
Notice that no comparison of the external {\em frequency} $\omega$ and
the DS {\em energy} $\varepsilon_s$ ever appears in the
Gross-Pitaevskii treatment.

Consider 1D Bose liquid subject to an external ac potential with
the wavenumber $q$ and frequency $\omega$.  Such a radiation may
be created using Bragg scattering technique~\cite{Bragg1,Bragg2}.
In these experiments  the ac potential has been created
by the interference pattern of two non-collinear optical beams with
the differential frequency $\omega$ and the $x$-component of the
differential wavevector $q$, [\onlinecite{Bragg_foot}].  At zero
temperature, according to the fluctuation-dissipation theorem, the
probability to absorb the radiation is given by the dynamic
structure factor (DSF), defined as the density-density correlation
function
\begin{equation}\label{S-definition}
    S(q,\omega)=\int \!\! dx dt\, e^{i(qx-\omega t)} \left\langle \rho(x,t)\rho(0,0) \right\rangle\, ,
\end{equation}
where $\rho(x,t)$ is the density operator.
Rewriting  DSF in the Lehman representation in terms of exact {\em many-body}
eigenstates of the system $|n\rangle$ with energies $\epsilon_n$, one finds
\begin{equation}\label{S-Lehman}
    S(q,\omega)=\sum\limits_n \left| \langle n|\rho_q|0\rangle
    \right|^2 \delta(\epsilon_n-\epsilon_0- \hbar\omega)\, ,
\end{equation}
where $\rho_q$ is the Fourier component of the density and $n=0$
corresponds to the ground state. Since the momentum is a good quantum
number, only the many-body states with the total momentum $p=\hbar q$
contribute to the sum in the r.h.s. of Eq.~(\ref{S-Lehman}).

\begin{figure}[h]
\includegraphics[width=0.9\columnwidth]{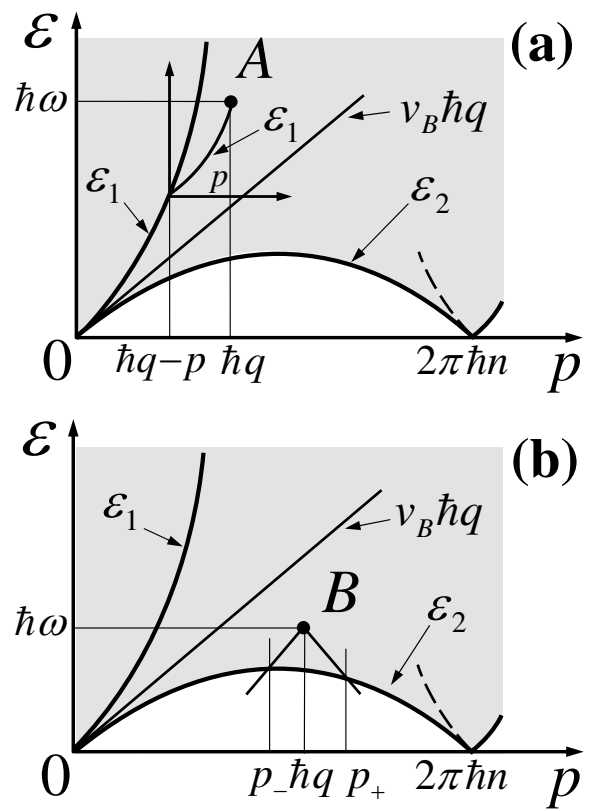}
\caption{
  Momentum--energy plane for excitations.  A photon is represented by
  a solid dot. (a) Absorption of a photon (dot marked by $A$) with
  energy and momentum slightly below the quasiparticle spectrum
  $\varepsilon_1(p)$, leads predominantly to creation of two
  quasiparticles with momenta $q-p$ and $p$ (the latter is determined
  by the shown geometrical construction). As $A$ approaches the line
  $\varepsilon= v_B p$, the number of excited quasiparticles
  increases. (b) Once photon energy-momentum (dot marked $B$) fall
  below the line $\varepsilon= v_B p$, its absorption involves creation
  of a soliton. Other excitations created in the course of absorption
  may be treated as phonons if $B$ is close to the boundary
  $\varepsilon_2(p)$. The shown
%
%
  ``sound cone'' with Bogoliubov velocity $v_B$ determines the range
  of possible momenta of a generated dark soliton, $p_-<p_s<p_+$.
  \label{fig:continuum}}
\end{figure}

For the model with the short-range repulsive interactions the
many-body spectrum has been evaluated exactly using the Bethe
ansatz (BA) method \cite{Lieb}.  Lieb  has identified two
characteristic modes in the excitation spectrum of the
model\cite{Lieb}, known as Lieb I and II modes with the dispersion
relations $\varepsilon_{1,2}(p)$, Fig.~\ref{fig:continuum}. The
two are given correspondingly by the particle and hole excitations
in the set of the BA quasi-momenta. The hole-like mode
$\varepsilon_2(p)$ is shown to be the lower bound of the many-body
spectrum with a given momentum $p$. According to
Eq.~(\ref{S-Lehman}) absorption is only possible if $\hbar\omega >
\varepsilon_2(\hbar q)$. Employing a numerical implementation of
the algebraic BA \cite{Slavnov}, Caux and Calabrese \cite{Caux06}
have shown that DSF is indeed non-zero for all energies in excess
of $\varepsilon_2(\hbar q)$ and is peaked at the particle-like
mode $\hbar\omega = \varepsilon_1(\hbar q)$. In the limit of the
weakly interacting gas the latter approaches the Bogoliubov
dispersion relation \cite{Lieb,Pitaevskii}
\begin{equation}
\label{Bogoliubov}
    \varepsilon_1(\hbar q)\to \hbar\omega_B(q) =v_B\hbar q\sqrt{1+(\hbar q/2mv_B)^2}\, ,
\end{equation}
where $v_B$ is the Bogoliubov sound velocity and $m$ is the boson
mass.

This observation offers a way to interpret absorption in a vicinity
of the Lieb I mode, $\varepsilon_1(p)$, in terms of weakly
interacting Bogoliubov quasiparticles.  Consider, {\it e.g.}, a
photon with some wave vector $q$ and energy $\hbar\omega$ slightly
below the value $\varepsilon_1(\hbar q)$, {\it i.e.},
$\hbar\omega\lesssim\varepsilon_1(\hbar q)$, see point A in
Fig.~\ref{fig:continuum}(a). The energy and momentum conservation
laws allow for such photon to create two Bogoliubov quasiparticles,
$\hbar\omega=\varepsilon_1(\hbar q -p)+\varepsilon_1(p)$. In the
limit $\varepsilon_1(\hbar q)-\hbar\omega\ll mv_B^2$, one of the two
particles has small momentum and may be viewed as a ``soft'' phonon.
For smaller initial photon energies, the resulting two
quasiparticles split the photon momentum more evenly, until the
photon energy reaches the limiting value
$\hbar\omega=2\varepsilon_1(\hbar q/2)$. If the photon energy is
decreased below this threshold, a creation of more than two
quasiparticles is needed to satisfy the conservation laws. Upon
further lowering $\hbar\omega$, more quasiparticles are created in
the process of photon absorption. Once the photon energy
$\hbar\omega$ approaches the line $\varepsilon=v_B\hbar q$,
Fig.~\ref{fig:continuum}, the energy and momentum of the absorbed
photon is split between infinitely many soft phonons.

Below this line the described process of dividing the energy and
momentum between the quasiparticles
does not work any more. Nevertheless the many-body spectrum persists
down to the lower value $\varepsilon_2(\hbar q) < v_B\hbar q$ and the
algebraic BA calculations \cite{Caux06} show that there is a
finite absorption probability in the energy window
\begin{equation}
\label{window}
    \varepsilon_2(\hbar q) < \hbar \omega <v_B\hbar q\, .
\end{equation}
What is the absorption mechanism in this window, where the
conservation laws forbid excitation of any number of quasiparticles
or phonons?

The clue to answer this question appeared in the 1976 paper of  Kulish,
Manakov and Faddeev \cite{Kulish}, who noticed that the hole-like Lieb II mode
approaches  dispersion relation of DS in the weakly interacting limit
\begin{equation}\label{LiebII}
    \varepsilon_2(p)\to \varepsilon_s(p)\, .
\end{equation}
It means that the many-body states with the energy in the vicinity
of $\varepsilon_2$ must be viewed as quantized DS particles.
Correspondingly the photon absorption in the energy window
(\ref{window}) necessarily involves excitation of DS along with
Bogoliubov quasiparticles and/or phonons. Consider e.g. a photon
with the energy immediately above the Lieb II mode, $\varepsilon_2
<\hbar\omega\lesssim \varepsilon_2+mv_B^2$ (point B in
Fig.~\ref{fig:continuum}~(b)). Drawing the ``sound cone'' with the
slope $v_B$ down to the intersections with $\varepsilon_2(p)$, one
finds the range of the possible momenta of DS
\begin{equation}\label{momentum-window}
    p_-<p_s<p_+\, ,\quad\quad p_\mp=p_\mp(q,\omega)
\end{equation}
which satisfy the conservation laws. Indeed, DS with the momentum
$p_-$ accompanied by a phonon, propagating in the direction of the
external momentum $\hbar q$, obviously satisfies the energy and
momentum conservation. Similarly, DS with the momentum $p_+$ must be
accompanied by the counter-propagating phonon. Any other soliton from
the momentum window (\ref{momentum-window}) requires excitation of a
certain superposition of the forward and backward propagating phonons.

In this paper we evaluate probability $W_{q\omega}(p_s)$ to excite DS
with the momentum $p_s$ in the range (\ref{momentum-window}) upon
absorption of a photon with the wavenumber $q$ and frequency $\omega
\geq \varepsilon_2(\hbar q)/\hbar$. We show that such a probability is
heavily shifted towards the lower boundary of the interval $p_s=
p_-(q,\omega)$, i.e. DS is preferentially excited along with the {\em
  forward} moving phonon.  At larger photon energies, while still in the
interval (\ref{window}), DS is excited with highest probability
along with the forward moving Bogoliubov quasiparticle.
%
%
Its energy and momentum may be found geometrically by plotting the
replica of the Bogoliubov dispersion curve which starts at some
point along the Lieb II mode, $\varepsilon_2(p)$, and passes through
the point $(\hbar q, \hbar\omega)$ representing external photon. We
also show that the total probability to excite any DS scales as a
power of the blue detuning from the energy threshold, $\int dp_s
W_{q\omega}(p_s)\propto (\hbar\omega-\varepsilon_2)^{\mu_2}$. The
exponent $\mu_2=\mu_2(q)$ is a function of photon wavenumber and the
strength of interactions between the bosons. In the relevant limit
of the weakly interacting gas, the exponent is large $\mu_2\gg 1$,
signifying the relative smallness of the photo-solitonic effect. As
we explain below, such a smallness is associated with the quantum
orthogonality catastrophe phenomenon~\cite{orthogonality}.

The rest of this paper is organized as follows. In section \ref{sec2}
we reproduce a derivation of DS solution of the Gross-Pitaevskii
equation to introduce notations and terminology. In section \ref{sec3}
we evaluate the probability to excite a specific DS upon absorption of
a photon. Section \ref{secDSF} is devoted to evaluation and discussion
of DSF i.e. the total photon absorption rate, resulting in DS
formation. Finally in section \ref{sec4} we discuss ways to observe
the effect experimentally along with the limitations of our theory.

\section{Dark Solitons}
\label{sec2}

To establish  notations let us briefly discuss the localized
solutions of the non-linear Gross-Pitaevskii
equation\cite{review1999}. Quasiclassically, this equation is
obeyed by the condensate wave function:
\begin{align}                                                                       \label{GP}
i  \partial_t \Psi + \frac{ 1 }{ 2 m }\, \partial_x^2 \Psi + c \left(
n  - |\Psi|^2 \right)\Psi = 0 \, ,
\end{align}
where $n=N/L$ is the average concentration and $L$ is the length of
the system. Hereinafter we switch to the units  with $\hbar=1$.
The interaction strength $c$ determines~\cite{Lieb} the
dimensionless parameter $\gamma = mc/n$ whose smallness
$\gamma\ll 1$ is the criterion of the weak interaction.

\begin{figure}[h]
\includegraphics[width=1.0\columnwidth]{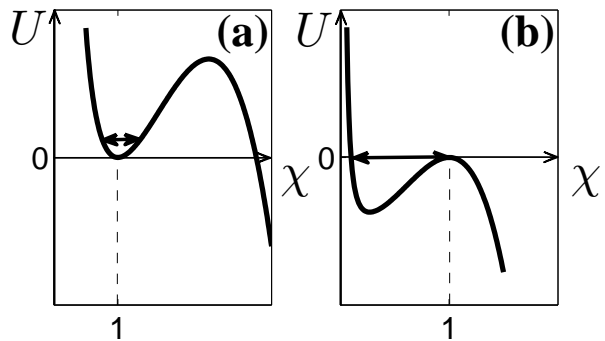}
 \caption{Effective potential $U(\chi)$ as
   given by Eq.~\eqref{potential} for (a) $v_S> v_B$; (b) $v_S < v_B$.
   The double-arrowed line designates the interval of variation of the
   normalized amplitude $\chi$ for physically allowed solutions of
   Eq.~\eqref{amplitude}.\label{fig:EffectPot}}
\end{figure}

Looking for a localized solution traveling with a certain velocity
$v_s$, one substitutes
\begin{equation}\label{soliton-form}
    \Psi(x,t)=\Psi_s(x-v_st)=\sqrt{n}\,\chi\, e^{i\vartheta}\,
\end{equation}
in  Eq.~(\ref{GP}) and finds two equations for the phase
$\vartheta(\xi)$ and the normalized amplitude $\chi(\xi)$, which are
functions of $\xi=x-v_st$.  The first of these equations acquires the
form of the continuity relation
\begin{equation}\label{cont}
\left[\chi^2\left(\vartheta\,'- mv_s\right)\right]'=0\,,
\end{equation}
where primes denote derivatives with respect to $\xi$. Using the fact
that far from the soliton $\chi(\pm\infty)=1$ and
$\vartheta\,'(\pm\infty)=0$, one finds $ \vartheta\,'=
mv_s(1-1/\chi^2)$.  Employing this relation, the equation for the
amplitude may be written in the form
\begin{equation}\label{amplitude}
    \chi\,''= - \frac{\partial U(\chi)}{\partial\chi}\, ,
\end{equation}
where the effective potential $U(\chi)$,  see
Fig.~\ref{fig:EffectPot}, is given by
\begin{equation}\label{potential}
    U(\chi)={m^2v_s^2\over 2} \left({1\over \chi^2} - {v_B^2\over v_s^2}\right)\left(1 - \chi^2\right)^2\,,
\end{equation}
with the Bogoliubov velocity $v_B=\sqrt{cn/m}$.

For $v_s>v_B$ the potential has the minimum at $\chi=1$ and the only
physically acceptable solutions of Eq.~(\ref{amplitude}) are small
oscillations around this minimum.  In a vicinity of $\chi=1$ the
potential (\ref{potential}) may be approximated as $U\approx
2m^2(v_s^2-v_B^2)(1-\chi)^2$ and therefore the small oscillation
solutions have the form $\chi-1\sim \cos(q(x-v_st))$ with
$q=2m\sqrt{v_s^2-v_B^2}$. Rewriting the last expression as
$v_s=v_B\sqrt{1+(q/2mv_B)^2}$, one may recognize it as the phase
velocity of the Bogoliubov mode. Correspondingly, the oscillation
frequency $qv_s=\omega_B(q)$ coincides with Eq.~(\ref{Bogoliubov}).
We thus conclude that the only solutions of GP equation which travel
with a supersonic velocity are Bogoliubov quasiparticles.

The situation is more interesting for $v_s<v_B$. In this case the
potential (\ref{potential}) exhibits a maximum at $\chi=1$, a minimum
at a smaller amplitude and a turning point at $\chi=v_s/v_B<1 $. The
solution with the proper boundary conditions, $\chi(\pm\infty)=1$, is
a trajectory which stays at the maximum and then exhibits a bounce
down to the turning point and back to the maximum. This is the DS
solution. To find it analytically, one may notice that
Eq.~(\ref{amplitude}) admits an integral of motion which for DS
solution reads as
\begin{equation}\label{integral-of-motion}
    {1\over 2}\, (\chi')^2 + U(\chi) = 0\, .
\end{equation}
Integrating this equation, one finds~\cite{Tsuzuki1970} for the wave
function (\ref{soliton-form})
\begin{align}
                                              \label{solitonGP}
\Psi_s =  \sqrt{n} \left[\cos \frac{ \theta_s }{ 2 } -
i \sin  \frac{ \theta_s }{ 2 }\, \tanh \left( \frac{ x - v_s
t}{l_s} \right) \right],
\end{align}
where
\begin{equation}\label{theta-s}
     \cos (\theta_s/2) =v_s/v_B
\end{equation}
with $\theta_s$ being the change of phase of the wave function across
the soliton. The soliton length $l_s$ is given by
\begin{equation}\label{l-s}
l_s^{-1} =  m v_B \sin (\theta_s / 2) = m\sqrt{v_B^2-v_s^2} \,.
\end{equation}

\begin{figure}[h]
\includegraphics[width=1.0\columnwidth]{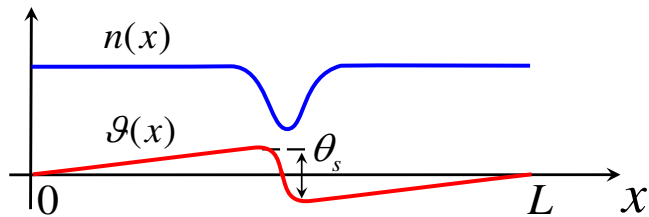}
\caption{(color online) Density $n(x)$ and phase $\vartheta(x)$
  profiles of a soliton in a system of length $L$. Note that the density perturbation is local,
  while the perturbation of phase is not.\label{fig:profile}}
\end{figure}
%

The number of particles pushed away from
the soliton core is
\begin{equation}\label{Ns}
N_s = \int\!\! dx \left(n-|\Psi_s|^2\right) = \frac{ 2K} { \pi}\, \sin { \theta_s \over 2 }\, ,
\end{equation}
where the ``quantum parameter'' $K=\pi n /(m v_B)$ depends on the
inter-particle interaction strength via the thermodynamic
compressibility which defines the velocity $v_B$. Notice that the
particle number may be very large, $N_s\gg 1$, in the limit of the
weakly interacting gas ($K\gg 1$).  The energy of the soliton is
given by
\begin{equation}\label{s-energy}
\varepsilon_s = \int\!\! dx \left[{1\over 2m}|\partial_x \Psi_s|^2  +{c\over 2}\left(n-|\Psi_s|\right)^2  \right]
= {4n v_B\over 3}   \sin^3 {\theta_s \over 2}\,.
\end{equation}
The calculation of DS momentum requires some care. The soliton {\em
  core} momentum, defined by the wave function
(\ref{solitonGP}) is
\begin{equation}\label{p-s-core}
    p_{sc}= \mbox{Im}\int\!\! dx \Psi_s^*\partial_x \Psi=  - n \sin \theta_s \,.
\end{equation}
However, one should take into account the periodic boundary conditions
which ensure that DS phase shift $\theta_s$ is uniformly spread over
the length of the entire system $L$, Fig.~\ref{fig:profile}. Although
this does not change the energy of the system in the thermodynamic
limit (indeed the corresponding contribution to the energy scales as
$n\theta_s^2/(mL)$), it produces a finite contribution $n\theta_s$ to
the momentum. As a result the total (core plus the rest of the
condensate) momentum of the DS state is
 \begin{equation}\label{p-s-total}
    p_{s}= p_{sc}+n\theta_s =   n\left(\theta_s- \sin \theta_s\right)  \,.
\end{equation}
Equations (\ref{s-energy}) and (\ref{p-s-total}) give an implicit
form of DS dispersion relation $\varepsilon_s(p_s)$.  The maximum of
the soliton energy corresponds to $\theta_s = \pi$, where both the
soliton velocity and core momentum vanish $v_s=p_{sc}=0$. The total
momentum, however, is finite $p_s=\pi n$ and is uniformly spread
across the entire condensate.  This is the true DS, in a sense that
the density vanishes in its center and the particle depletion
reaches its maximal value $N_s=2K/\pi$. Away from the point
$\theta_s =\pi$ soliton's velocity is finite $v_s\neq 0$ as well as
the density at any point. Because of the latter such solitons are
sometimes called {\em
  grey}. Their velocity approaches sound velocity $v_B$ when the total
momentum approaches zero or $2\pi n$, while the energy and $N_s$
both decrease. Clearly the concept of the classical soliton looses
sense when the number of particles pushed away from the core is
comparable to one, $N_s\lesssim 1$. This takes place when
$\theta_s\lesssim 1/K\ll 1$, and therefore at $|p_s|\lesssim n/K^3$,
and in intervals of the same width around the point $p=2\pi n$.

In the limit of the weak interactions $\gamma\ll 1$ (i.e. $K\gg
1$) the DS dispersion relation given by Eqs.~(\ref{s-energy}),
(\ref{p-s-total}), approaches the Lieb II mode, plotted in
Fig.~\ref{fig:continuum}. The convergence is not uniform and the
two significantly deviate from each other in the narrow intervals
of momenta near zero $|p_s|\lesssim nK^{-3/2}$ and similarly near
$2\pi n$ [\onlinecite{unpub}]. Notice that at the boundaries of
this interval the number of particles pushed away from the soliton
core is still large $N_s\approx \sqrt{K}>1$. It is this condition,
rather than the weaker one $N_s\gtrsim 1$, which determines the
validity of the soliton approach. We shall return to this
observation in section \ref{secDSF}.

\section{Excitation of Dark Solitons}
\label{sec3}

Consider a Bose gas subject to a weak space and time dependent
external potential $V_0\cos (qx-\omega t)$. According to the Golden
Rule (c.f. Eq.~(\ref{S-Lehman})), the system may absorb quanta of this
field if its many-body spectrum possesses excited states with the
momentum $ q$ and energy $\omega$.  It follows from the exactly
solvable model~\cite{Lieb} that such states form a continuum whose
energy is bound from below by the Lieb II mode $\varepsilon_2(q)$.  As
argued in the Introduction absorption of quanta with the energy in the
range given by Eq.~(\ref{window}) is associated with creation of DS
along with the phonons or quasiparticles.

To evaluate the probability of such a process it is convenient to
think of it in terms of the space-time evolution of a state
resulting from the photon absorption by the system initially in the
ground state.
To this end we notice that the photon absorption first creates a
{\em virtual} state of the condensate with a {\em local}
perturbation of the condensate wavefunction. Since the photon
carries momentum $q$ and no extra particles, so does the initial
local perturbation. Subsequently this perturbation evolves and
eventually  takes a form of a superposition of {\em real}
excitations, i.e. conserving overall energy $\omega$ in addition
to the momentum $q$. We expect that such a final state contains a
soliton with the momentum $p_s\approx q$ and core energy
$\varepsilon_s(p_s) < \omega$. The small excess energy
$\omega-\varepsilon_s>0$ is carried away by phonons, propagating
with the sound velocity $v_B$.

The initial separation of the soliton core from a bunch of phonons
takes a short time, which may be estimated as $\tau_s=l_s/v_B$. The
soliton core is the density depletion, which carries momentum $-
n\sin\theta_s$ (which is very different from $p_s\approx q$) and
$-N_s$ particles. At times $t>\tau_s$ the core propagates without
dispersion and behaves as a free particle with the energy
$\varepsilon_s$. The remaining momentum $q+n\sin\theta_s \approx
p_s+n\sin\theta_s =n\theta_s$, cf. Eq.~(\ref{p-s-total}), and $N_s$
particles, initially localized on a scale $\sim l_s$, must be
carried away and spread over the entire system at $t\gg\tau_s$ by
the phonons. As explained above, despite the fact that the phonons
must carry away large number of particles $N_s$ and large momentum
$n\theta_s$, their final energy is small, $\omega-\varepsilon_s\ll
v_Bn\theta_s$. Therefore this is the low-probability event, or the
``under-barrier'' process, which should be described as the
imaginary time evolution \cite{Landau-Lifshitz,LevitovShytov} of the
phonon system~\cite{Matveenko}.

To develop such a description we start from the imaginary time $\tau$
action for the interacting Bose field
\begin{equation}\label{action}
    S=\!\int\! d\tau dx \left[\bar \Psi\partial_\tau \Psi - {1\over 2m}|\partial_x \Psi|^2 +cn|\Psi|^2-{c\over 2}\, |\Psi|^4\right].
\end{equation}
It is convenient to parameterize the complex field as
$\Psi=\sqrt{n+(\partial_x \varphi/\pi) }\, e^{i\vartheta}$, where
$\varphi(x,\tau)$ and $\vartheta(x,\tau)$ are two real fields
describing density and phase fluctuations correspondingly. Assuming
small density fluctuations $\partial_x\varphi \ll\pi n$ and
linearizing the resulting action, one finds
\begin{equation}\label{action1}
    S=\!\int\! d\tau  \left[{i\over \pi}\int\! dx\, \partial_x\varphi\, \partial_\tau\vartheta - H_{sw} \right]\,,
\end{equation}
where the hydrodynamic Hamiltonian of the sound waves is given by
\cite{Haldane}
\begin{equation}                                             \label{Hhydro}
H_{sw} = \frac{ \,v_B}{2\pi}\int\!dx \left[K^{-1} (\dx{\varphi})^2 +
K(\dx{\vartheta})^2\right]\, .
\end{equation}
We have omitted terms $\sim(\partial_x^2 \varphi)^2$ in the
Hamiltonian, which is equivalent to restricting the spectrum of
Bogoliubov quasiparticles to the phonon branch only.  This
approximation is sufficient for treating photon absorption close to
the soliton threshold.

Taking the variations of the imaginary-time action over $\varphi$ and
$\vartheta$, one finds the semiclassical equations of motion
\begin{eqnarray}
\label{equation-of-motion}
  \partial^2_{\tau x}\varphi + i v_BK\partial_x^2 \vartheta &=& \pi N_s
  \left[\delta(x)\delta(\tau) -\delta(x-\bar x)\delta(\tau-\bar\tau)\right],    \nonumber \\
  \partial^2_{\tau x} \vartheta + i  {v_B\over K}\,\partial_x^2 \varphi &=& \theta_s
  \left[\delta(x)\delta(\tau) -\delta(x-\bar x)\delta(\tau-\bar\tau)\right].\nonumber
\end{eqnarray}
The right hand sides of these equations contain sources which describe
the feedback of DS creation at the point $x=\tau=0$ and its subsequent
destruction at the point $x=\bar x$, $\tau=\bar \tau$. As discussed
above, such creation (destruction) of DS is associated with
practically instantaneous and local injection (removal) of $N_s$
particles and momentum $n\theta_s$ into (out of) the phonon modes. The
sources on the r.h.s. of the equations of motion do just that. The
equations of motions are straightforwardly solved by the Fourier
transformation.  Substituting such a solution back into
Eq.~(\ref{action1}), one finds for the imaginary time action
 \begin{equation}\label{action2}
    S(\bar x,\bar\tau) = \mu_+ \ln\left(1+\frac{\bar x-iv_B\bar \tau}{il_s}\right) + \mu_- \ln\left(1+\frac{\bar x+iv_B\bar \tau}{il_s}\right),
 \end{equation}
where we introduced notations
\begin{equation}
\mu_{\pm} =  \frac{\left(K\theta_s  \pm \pi
N_s  \right)^2}{4\pi^2K} = \frac{K}{\pi^2} \left(\frac{\theta_s}{2} \pm
\, \sin \frac{ \theta_s}{ 2}  \right)^2 \label{exponents}
\end{equation}
and employed Eq.~(\ref{Ns}) in the last equality in the r.h.s. of
Eq.~(\ref{exponents}). Here $\theta_s$ is the parameter of a created
DS, it is related to the DS momentum $p_s\approx q$ through
Eq.~(\ref{p-s-total}). The soliton length $l_s$ appears in
Eq.~(\ref{action2}) as a short distance cutoff. Indeed, one should
understand that the actual spatial (temporal) extent of the
delta-functions on the r.h.s. of the equations of motions is the
soliton size $l_s$ ($\tau_s=l_s/v_B$).

To find a probability of creating DS with the momentum $p_s$ upon
absorbing a photon $(q,\omega)$, one needs to evaluate the Fourier
transform of the square of the semiclassical matrix element given by
$e^{-S}$. Specifically,
\begin{equation}\label{WFourier}
W_{q,\omega}(p_s)= \mbox{Re}\!\int\!\! \frac{d\bar x\,
d\bar{t}}{l_s}\,\,\, e^{-S(\bar x,\bar{t}) - i(q-p_s)\bar
x+i(\omega-\varepsilon_s)\bar{t}} ,
\end{equation}
where we took into account that the momentum $p_s$ and energy
$\varepsilon_s(p_s)$ are carried away by the soliton and therefore
should not be absorbed by the phonons. The analytical continuation
performed in Eq.~\eqref{WFourier} to the real frequencies $i \omega
\rightarrow \omega$ is accompanied by the time integration contour
(Wick) rotation $\bar{\tau} \rightarrow i \bar{t}$ to ensure
convergence.

\begin{figure}[h]
\includegraphics[width=0.9\columnwidth]{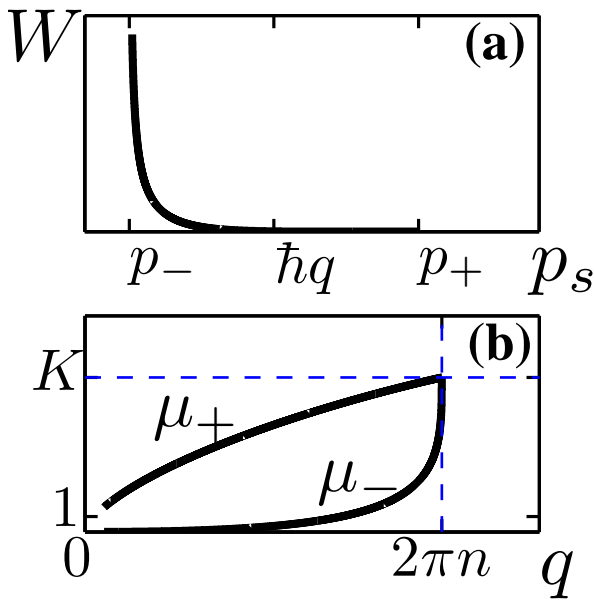}
\caption{ (a) Soliton creation rate as
  a function of the soliton momentum $p_s$. (b) Momentum dependence of the
  exponents $\mu_{\pm}$  for $K=10$.
\label{fig:crossection}}
\end{figure}

To evaluate the integral in Eq.~(\ref{WFourier}), we take into
account that for small energy excess $\omega-\varepsilon_s(q)\ll
mv_B^2$ the range of the allowed soliton momenta $p_s$ is rather
narrow, see Fig.~\ref{fig:continuum}~(b), and centered around the
photon momentum $q$. One may therefore expand the soliton energy
as $\varepsilon_s(p_s)\approx \varepsilon_s(q)+(p_s-q)v_s$ and
find for the boundaries of the possible soliton momenta, see
Eq.~(\ref{momentum-window}) and Fig.~\ref{fig:continuum}~(b),
\begin{align}                                                                       \label{interval}
 p_{\pm}(q,\omega) =
q \pm \frac{\omega - \varepsilon_s(q)}{v_B \pm v_s(q)}\,; \quad\,\,\,\, p_+ - p_- \ll p_s\,.
\end{align}
Adopting these notations and performing the straightforward
integrations in Eq.~(\ref{WFourier}), one finds
\begin{align}                                                                       \label{distribution}
W_{q,\omega}(p_s) \propto  \frac{l_s}{v_B}\left[\frac{p_+  -
p_s}{l_s^{-1}} \right]^{\mu_+ - 1}
\left[\frac{p_s  -  p_-}{l_s^{-1}}\right]^{\mu_- - 1}\, .
\end{align}
Therefore the soliton creation rate is characterized by the
power-law dependencies on the deviations of the soliton momentum
from the upper and lower kinematic boundaries $p_\pm(q,\omega)$,
Fig.~\ref{fig:crossection}~(a). The corresponding exponents
$(\mu_\pm-1)$, see Fig.~\ref{fig:crossection}~(b), are functions
of the soliton parameter $\theta_s=\theta_s(q)$ and the
quantum parameter $K$ as given by Eq.~(\ref{exponents}).  Since
$\mu_-<\mu_+$, the probability to excite the soliton is heavily
shifted towards the lower boundary $p_-$,
Fig.~\ref{fig:crossection}(a). I.e. the soliton is preferentially
accompanied by the {\em forward} moving phonons (in the direction
of the photon momentum $q$).

\section{Dynamic Structure Factor}
\label{secDSF}

Another quantity of interest is the total absorption rate of
photons with a given $q$ and $\omega\gtrsim\varepsilon_s(q)$,
which results in creation of a soliton with an unspecified
momentum. This quantity is nothing but DSF $S(q,\omega)$ of the 1D
Bose gas. Integrating $W_{q,\omega}(p_s)$,
Eq.~(\ref{distribution}), over the soliton momenta $p_s$, one
finds for DSF in an immediate vicinity of the lower spectral
boundary $\omega \gtrsim \varepsilon_{2}(q)$
\begin{align}                                                                       \label{eq:res}
S(q,\omega) = \int\limits_{p_-}^{p_+} \! dp_s\, W_{q,\omega}(p_s)\propto {1\over v_B}
\left(\frac{p_+ - p_-}{l_s^{-1}} \right)^{\mu_2} ,
\end{align}
where the exponent is given by $\mu_2 = \mu_+ + \mu_- - 1$.
According to Eq.~(\ref{exponents}), the exponent is expressed through the parameters of the soliton $\theta_s$,
which in turn is related to the soliton momentum through Eq.~(\ref{p-s-total}),
where  $p_s= q$. As a result
\begin{align}                                                                      \label{eq:exp}
\mu_2(q) =
\frac{2K}{\pi^2}\left[\left(\frac{\theta_s}{2}\right)^2
                     +\left(\sin\frac{\theta_s}{2}\right)^2
                \right]-1\,.
\end{align}
For the true DS $\theta_s=\pi$ and therefore $\mu_2(\pi n)\approx
0.70K$.
Notice that at $q\to 2\pi n$ Eq.~(\ref{eq:exp}) yields $\mu_2\approx
2K-1$, different from DSF exponent $K-1$ established in the
framework of the Luttinger liquid theory \cite{Haldane}. The latter
is applicable above the dashed line in Fig.~\ref{fig:continuum}.

Employing Eq.~(\ref{interval}), one may rewrite DSF (\ref{eq:res})
in the following form
\begin{align}                                                                       \label{DSF-res}
S(q,\omega) \propto {1\over v_B} \left[ \frac{ \omega -
\varepsilon_s(q)}{ \Lambda(q) }\right]^{\mu_2(q)}\theta\left(\omega -
\varepsilon_s(q)\right)\, ,
\end{align}
where the cutoff energy is
\begin{equation}\label{cutoff}
    \Lambda(q)=\frac{mv_B^2}{2}\, \sin^3\frac{\theta_s}{2}\approx \frac{\varepsilon_s(q)}{K}\,.
\end{equation}
Being multiplied by the intensity of the radiation $V_0^2$, DSF gives
a number of solitons excited per unit time and per unit length of the
irradiated 1D gas.

The power law behavior of DSF near the lower spectral boundary
$\varepsilon_2( q)\approx \varepsilon_s(q)$ was derived earlier by the
present authors and M.~Pustilnik in Ref.~[\onlinecite{bosons2007}].  There a
mapping between 1d Bose and Fermi systems was used to prove the
presence of the power law non-analyticity and evaluate the exponent
$\mu_2(q)$. However, the method adopted there allowed us to deduce the
exponent only in the limit of strongly interacting bosons $\gamma\gg 1$
(since the latter is mapped onto weakly interacting fermions, treated
in Ref.~[\onlinecite{Khodas2006}]).  Later a method to extract the edge
exponent $\mu_2$ for an arbitrary interaction parameter from the BA
solution was suggested in
Refs.~[\onlinecite{Imambekov,Affleck-Pustilnik,Khodas-unp}].

\begin{figure}[h]
\includegraphics[width=0.9\columnwidth]{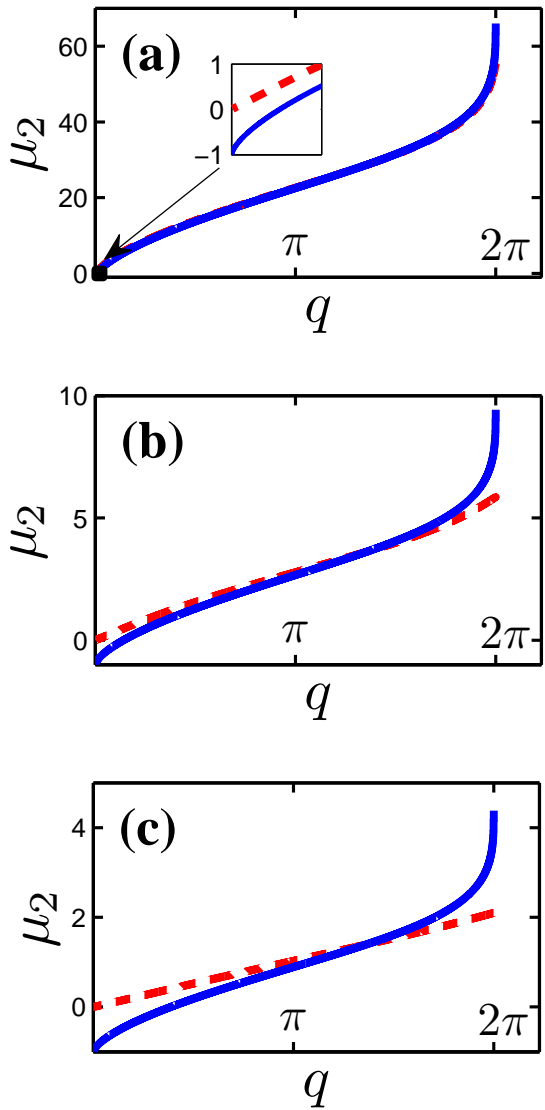}
\caption{(color online) Edge exponent $\mu_2(q)$, cf.
Eq.~(\ref{DSF-res}), as a function of momentum in units of $n$ for (a)
$\gamma = 0.05,\, K=33$;  (b) $\gamma = 0.4,\,  K = 5.2$; (c)
$\gamma = 1,\,  K =2.7$. The full (blue) line is the semiclassical
result (\ref{eq:exp}); dashed (red) line is the Bethe ansatz
solution of Ref.~[\onlinecite{Imambekov}]. \label{fig:Bethe}}
\end{figure}

Fig.~\ref{fig:Bethe} shows comparison between the semiclassical result
Eq.~(\ref{eq:exp}) and the numerical solution of BA equations
\cite{Imambekov} for the edge exponent $\mu_2$ as a function of the
momentum (in units of $n$).  The agreement between the two approaches
becomes progressively better for weaker interactions (the only limit
where the soliton picture holds, see Introduction). Such an agreement
suggests that the interpretation of the photon absorption near the
lower spectral edge as a formation of solitons is indeed consistent
with a fully quantum many-body calculation. The latter
\cite{Imambekov} does not rely on existence of solitons at all. We
consider it as a strong confirmation of the thesis that absorption of
an ac quanta in the frequency window (\ref{window}) results in the
formation of DS.

Notice that even in the weakly interacting limit, the
semiclassical prediction (\ref{eq:exp}) deviates from the exact
one at very small momenta, see the inset in
Fig.~\ref{fig:Bethe}(a). This is to be expected, since as was
discussed in section \ref{sec2}, the soliton picture looses its
validity at sufficiently small momenta.  Inspecting
Eq.~(\ref{eq:exp}), one notices that the semiclassical exponent
becomes negative at $\theta_s\lesssim\pi/\sqrt{K}$, contrary to
the exact results, Fig.~\ref{fig:Bethe}.  Using Eq.~(\ref{Ns}),
one finds that this corresponds to the number of missing particles
in the soliton core $N_s\approx \sqrt{K}$.  This observation
collaborates with the discussion presented in the end of section
\ref{sec2}, which suggests that the semiclassical treatment looses
validity for the very grey solitons with $N_s<\sqrt{K}$, i.e.
$|p_s|<nK^{-3/2}$, cf. Eq.~(\ref{p-s-total}).  We stress that the
power-law behavior of DSF at the exact lower spectral boundary
$\varepsilon_2(q)$ is valid for any momentum. However for
$q<nK^{-3/2}$ the semiclassical approximation for the exponent
$\mu_2(q)$ fails. Instead, the exact exponent
\cite{bosons2007,Imambekov} scales linearly with momentum
$\mu_2(q)\sim K^{3/2}q/n$.

\section{Discussion}
\label{sec4}

We have shown that the absorption of a photon with the energy above a
certain threshold leads to formation of the DS.  In the narrow energy
window $\varepsilon_s<\omega<\varepsilon_s+\Lambda$ the total soliton
formation {\em rate per unit length} of 1D Bose cloud is given by
$S(q,\omega)V_0^2$, where DSF is given by Eq.~(\ref{DSF-res}) and
$V_0\cos(\omega t-qt)$ is the external ac potential applied to 1D Bose
gas.  The momentum-resolved rate is given by Eq.~(\ref{distribution}),
{\it i.e.}, solitons with the momentum in the interval $p_s\pm
dp_s/2$, where $p_s$ belongs to the window (\ref{momentum-window}),
are created with the rate $W_{q,\omega}(p_s)V_0^2dp_s$.

An important question is what are the corresponding rates for a larger
energy of the photon: $\varepsilon_s+\Lambda<\omega<v_Bq$. According
to the arguments given in the Introduction, absorption of such a
photon should necessarily lead to DS formation. Yet our calculations
are not directly applicable in this case. Indeed, we have used
linearized dispersion relation for the quasiparticles (phonons)
excited along with DS. For energies above $\varepsilon_s+\Lambda$ such
an approximation is not valid. This is because Bogoliubov
quasiparticles with momenta above $mv_B$ that take the excess energy
can not be well approximated by phonons.
The photon absorption is dominated by creation of a DS and single
quasiparticle moving
in the direction of the wavevector $q$ (i.e. moving forward). Thus
the parameters of the typical DS may be found by plotting a
replica of the Bogoliubov spectra which starts at some point along
the absorption edge $\varepsilon=\varepsilon_s(q)$ and passes
through $(q,\omega)$.  The starting point prescribes DS momentum
and energy. We expect that the power-law Eq.~(\ref{DSF-res})
saturates at an excess energy of order $\Lambda(q)$, i.e. at
$\omega-\varepsilon_2\approx  \Lambda$.  As a result, DS formation
rate per unit length may be estimated as $(V_0^2/\hbar^2
v_B)e^{-\alpha \mu_2(q)}$, with a numerical factor $\alpha\approx
1$.

Recalling that for a true DS $\mu_2=0.7K= 1.1 N_s$, one realizes
that the photo-solitonic rate is exponentially suppressed with the
increase of DS depleted  particle number $N_s$. Physically the
origin of this smallness is in the orthogonality phenomenon: the
state of the system immediately after absorption of the photon is
almost orthogonal to the state with the soliton causing a
re-distribution of density and phase of the condensate in the
one-dimensional system. The corresponding matrix element is
exponentially small in the parameter $N_s$. This fact dictates a
rather stringent limitations on the experimental observability of
the photo-solitonic effect.  Increasing interactions (i.e.
decreasing $K$) makes the exponential factor in the photo-solitonic
rate less severe, on the other hand it simultaneously decreases
$N_s$, making it more difficult to observe the excited solitons.
Assuming $V_0/(2 \pi \hbar) = 100\,\mathrm{Hz}$,
Ref.~[\onlinecite{Bragg2}] we estimate the soliton production rate
for the relatively ``light'' solitons with $N_s=10$ as $1$ event per
Bragg pulse of duration of $2$ seconds.

\begin{acknowledgments}
  We thank A. Abanov, J.-S. Caux, D. Gangardt, D. Gutman, V.~Gurarie,
  and A.~Imambekov for numerous discussions.  This research is supported by
DOE Grant No. DE-FG02-08ER46482 and A.P. Sloan foundation.
\end{acknowledgments}

\end{document}